\documentclass[floats, prd, eqnum, showpacs, nofootinbib, 
twocolumn, 
dvipdfmx, 
eqsecnum]{revtex4-1}

\usepackage{color,graphicx}
\usepackage{amsfonts}
\usepackage{amssymb}
\usepackage{comment}
\usepackage{ulem}
\usepackage{hyperref}
\hypersetup{
 setpagesize=false,
 bookmarksnumbered=true,%
 bookmarksopen=true,%
 colorlinks=true,%
}

\begin{document}

\title{Neutral particle collisions near Gibbons-Maeda-Garfinkle-Horowitz-Strominger black holes after shadow observations}
\author{Naoki Tsukamoto}\email{tsukamoto@rikkyo.ac.jp}
\author{Ryotaro Kase}

\affiliation{
Department of Physics, Faculty of Science, Tokyo University of Science, 1-3, Kagurazaka, Shinjuku-ku, Tokyo 162-8601, Japan }

\begin{abstract}
A Gibbons-Maeda-Garfinkle-Horowitz-Strominger (GMGHS) black hole with a magnetic charge (or an electric charge) has noteworthy features 
that its scalar curvature near the event horizon of the black hole with the almost maximal charge can be extremely large. 
The large curvature, which is related with the gravity on a finite-sized object or between two points, causes high center-of-mass energy for two neutral particles 
near the almost maximally charged GMGHS black hole. Recently, the Event Horizon Telescope Collaboration gave the bound on the charge of black holes 
from the shadow and mass observations of black holes under an assumption that the diameter of observed rings are proportion to that of photon spheres. The photon sphere would be less related with the curvature, 
since it is determined by the behavior of one photon or one ray neither two photons nor two rays. Thus, the high-energy neutral particle collision 
and the black hole shadow observations would be complementary to distinguish the GMGHS black hole from other black hole solutions. 
In this paper, we investigate a new way to compare the center-of-mass energy for neutral particle collisions in the GMGHS spacetime 
and other black hole spacetimes. 
From the shadow observations and the mass observations under the assumptions on the effect of black hole charges, 
we can put constraints on the center-of-mass energy of the particles.  We apply our method to shadow and mass observations of M87* and Sagittarius~A*.  
We find that the center-of-mass energy of neutral particles near the GMGHS black holes cannot be extremely large under the observational constraints, and conclude that the GMGHS spacetimes are hardly distinguishable from the Reissner-Nordstr\"{o}m spacetimes by the particle collisions 
if we apply the shadow and mass observations at $1 \sigma$ probability.
\end{abstract}

\date{\today}

\maketitle

\section{Introduction}
Black holes are well-known compact objects predicted in general relativity. 
The first direct detection of gravitational waves emitted by a black hole binary has been reported by LIGO Scientific Collaboration and Virgo Collaboration~\cite{LIGOScientific:2016aoc}
and the shadow images of supermassive black hole candidates at M87* and Sagittarius~A* (Sgr~A*) have been reported by Event Horizon Telescope Collaboration~\cite{EventHorizonTelescope:2019dse,EventHorizonTelescope:2022wkp}.    
The investigation of phenomena in a strong gravitational field near the black holes will be important to understand black hole spacetimes and to interpret the black hole observations.  

In 2009, Ba\~{n}ados, Silk, and West
studied two neutral particles colliding around a Kerr black hole and they found that the center-of-mass energy of the particles can be arbitrarily high in a near horizon limit if the Kerr black hole has an extreme event horizon and if either one of the two particles has the critical angular momentum~\cite{Banados:2009pr}.
The collision of neutral particles with the infinite center-of-mass energy in the extremal rotating Kerr spacetime is called the Ba\~{n}ados-Silk-West (BSW) process.
The BSW process has been strictly examined in the aspects of 
the existence of extreme Kerr black hole~\cite{Berti:2009bk}, 
gravitational radiation and its backreaction of the process~\cite{Berti:2009bk}, 
an infinite proper time of a falling particle~\cite{Jacobson:2009zg}, and 
self-gravity of falling objects~\cite{Kimura:2010qy,Ogasawara:2018gni}.
The BSW process of an equatorial plane~\cite{Harada:2011xz} and the BSW collision at the innermost stable circular orbit~\cite{Harada:2010yv} were also investigated.
\footnote{ 
In Refs.~\cite{Piran:1975, Piran:1977dm}, Piran, Shaham, and Katz 
investigated neutral particles colliding around a Kerr black hole. They pointed out that the center-of-mass energy of the head-on collisions of the particles in the non-extremal Kerr black hole spacetime is unbound on the horizon but that the high-energy collisions are hard to occur astrophysically. See Ref.~\cite{Ovcharenko:2024xkk} for the details of the head-on collision near the horizons.
}

In 2010, the electromagnetic counterpart of BSW collision was found by Zaslavskii~\cite{Zaslavskii:2010aw}. 
In the Zaslavskii process, the center-of-mass energy of the charged particles can be arbitrarily high in a near horizon limit 
around an extreme charged Reissner-Nordstr\"{o}m (RN) black hole if the either one of the two particles has a critical electric charge. 
Since the Zaslavskii process is easier to handle than the BSW process, a number of studies have been done such as 
the electromagnetic counterpart of a collisional Penrose process~\cite{Penrose:1969pc,Denardo:1973pyo} after the Zaslavskii collision~\cite{Zaslavskii:2012ax,Nemoto:2012cq,Tsukamoto:2022gxx}, the collisions of thin shells~\cite{Kimura:2010qy,Nakao:2017xwe}, 
multiple process~\cite{Kokubu:2020jvd,Kokubu:2021cwj} and a higher-dimensional Zaslavskii process~\cite{Tsukamoto:2013dna}.
 
In the BSW and Zaslavskii processes, the particle with the critical angular momentum or the critical charge almost stops in a radial direction near outside the extremal horizon while the other particle with a subcritical angular momentum or a subcritical charge moves with an almost speed of light there. 
Consequently, the relative velocity of the colliding particles reaches a speed of light on the extremal event horizon causing the unbounded center-of-mass energy~\cite{Jacobson:2009zg,Zaslavskii:2011dz,Harada:2014vka}. 
The unbounded center-of-mass energy in the BSW and Zaslavskii processes are considered universal property of a black hole spacetime 
with the extremal event horizon~\cite{Zaslavskii:2010jd}. 
Several high-energy particle collisions, which are different from the BSW and Zaslavskii processes, were also found, e.g.,
particle collisions near wormholes~\cite{Tsukamoto:2014swa,Tsukamoto:2015hta,Zaslavskii:2018kix,Krasnikov:2018nga,Tsukamoto:2019ihj},
naked singularity~\cite{Patil:2015fua}, a conical singularity~\cite{Abdujabbarov:2013qka}\footnote{In Ref.~\cite{Abdujabbarov:2013qka}, Abdujabbarov~\textit{et al.} claimed that infinite center-of-mass energy of two particles without the critical angular momentum occurs in a 5-dimensional extremal Kerr spacetime without infinite scalar curvatures. However, it was pointed out that this unbound center-of-mass energy is caused by the conical singularity on the extremal event horizon~\cite{Tsukamoto:2013dna}.}.    

In Einstein-Maxwell-dilaton theories in which the dilaton field directly coupled to the electromagnetic field, 
there is an exact black hole solution known by the name of Gibbons-Maeda-Garfinkle-Horowitz-Strominger (GMGHS) solution. 
This solution was first discovered in Ref.~\cite{Gibbons:1982ih,Gibbons:1987ps} and it describes electrically and/or magnetically charged 
black hole with a nontrivial dilation field in a static and spherically symmetric spacetime. 
The same solution without the electric charge was rediscovered in Ref.~\cite{Garfinkle:1990qj}. 
Uniqueness of the GMGHS black hole with the pure electric charge was proved by Masood-ul-Alam~\cite{Masood-ul-Alam:1993thx}. 
The uniqueness was extended to the GMGHS black hole including the magnetic charge by Mars and Simon~\cite{Mars:2001pz}
and the new proof of their uniqueness was given by Nozawa~\textit{et al.} in Ref.~\cite{Nozawa:2018kfk}. 
Kase and Tsujikawa investigated the linear stability of black holes in Maxwell-Horndeski theories and 
confirmed that the GMGHS black hole without the magnetic charge is stable~\cite{Kase:2023kvq}. 
Pope~\textit{et al.} prove the linear stability of the GMGHS black hole without the magnetic charge~\cite{Pope:2024ncb}.
The GMGHS solution was extended to the case of arbitrary number of gauge fields~\cite{Deshpande:2024itz}.

Since a purely electrically charged GMGHS black hole can be obtained by applying the electromagnetic duality transformation \cite{Gregory:1992kr,Horne:1992bi} to the purely magnetically charged GMGHS black hole, the motions of neutral particles or photons in these black hole spacetimes are the same. 
Several aspects of the motions of the particles and the photons were investigated well both theoretically 
and observationally, e.g., 
the innermost stable circular orbit~\cite{Maki:1992up,Pradhan:2012id},
shadow image or lensing ring images~\cite{Bhadra:2003zs,Hioki:2008zw,Heydari-Fard:2021pjc,Wielgus:2021peu,Kocherlakota:2022jnz,daSilva:2023jxa,Claros:2024atw},
null geodesics~\cite{Fernando:2011ki},
accretion disks~\cite{Karimov:2018whx},
accretion fluid~\cite{Bahamonde:2015uwa},
pulsar light curves~\cite{Sotani:2017bho}, and
the motion of the S2 star~\cite{DeLaurentis:2017dny,Fernandez:2023kro}. 
The constraints on either the electric charge or the magnetic charge of M87*~\cite{EventHorizonTelescope:2021dqv} and Sgr~A*~\cite{EventHorizonTelescope:2022xqj,Vagnozzi:2022moj} were also studied by using EHT shadow observation and mass observations.
On the other hand, we would distinguish the black holes by phenomena related to dynamics of the spacetimes, perturbation of the spacetimes, and the electromagnetic forces from the black holes to charged particles, do depend on the details of the dilaton and other scalar fields and electromagnetic fields. 
For instance, gravitational waves~\cite{Hirschmann:2017psw,Julie:2018lfp}, the stability of black holes~\cite{Gannouji:2021oqz,Kase:2023kvq}, and the Blandford-Znajek mechanism~\cite{Banerjee:2020ubc,Chatterjee:2023wti}, correspond to such studies. 

Pradhan~\cite{Pradhan:2013hxa} and Fernando~\cite{Fernando:2013qba} found that 
the center-of-mass energy for neutral particle collisions around 
a GMGHS black hole with the magnetic charge can be unboundedly high.
The mechanism is very different from the BSW and Zaslavskii processes.
The high center-of-mass energy for the neutral particle collision around the GMGHS black hole is caused by its noteworthy features 
that its scalar curvature near the event horizon of the black hole with the almost maximal charge can be extremely large. 
Zaslavskii~\cite{Zaslavskii:2016bic} generalized the particle collision with the unbounded center-of-mass energy and classified particle collisions with arbitrary high center-of-mass energy near the GMGHS black hole~\cite{Pradhan:2013hxa,Fernando:2013qba},
near a black hole in Brans-Dicke theory~\cite{Kim:1998hc} investigated by Sultana and Bose~\cite{Sultana:2015avz}, and
near naked singularities in a Fisher spacetime or a Janis-Newman-Winicour spacetime~\cite{Fisher:1948yn,Bergmann:1957zza,Buchdahl:1959nk,Janis:1968zz,Ellis:1973yv,Bronnikov:1973fh,Wyman:1981bd,Martinez:2020hjm} 
investigated by Patil and Joshi~\cite{Patil:2011aa}, 
into the same class.

Recently, general-relativistic magnetohydrodynamical (GRMHD) simulations in a black hole spacetime with the same metric as GMGHS black hole with only one of the magnetic charge or the electric charge were performed in Ref.~\cite{Mizuno:2018lxz} and the difficultly to distinguish the GMGHS black hole from the Kerr black hole was pointed out~\cite{Mizuno:2018lxz,Roder:2023oqa}. 
In this work, we investigate whether it is possible to distinguish 
the GMGHS black hole with only one of the magnetic charge or the electric charge 
from a RN black hole 
by using the center-of-mass energy for neutral particle collisions around the black holes taking 
the black hole shadow and mass observations into account.
We construct the method to constrain the center-of-mass energy from the shadow and the mass observations 
under assumptions on effect of black hole charges, and apply it to the observations of M87* and Sgr~A*.
We concentrate on the neutral particle collision since it is not disturbed by external magnetic fields in astrophysical situations 
and our result is also valid in other black hole spacetimes with the same metric as the GMGHS spacetime.

This paper is organized as follows.  
We investigate neutral particle motion and collision in Sec.~II.
We consider the Schwarzschild, GMGHS, and RN black holes in Sec.~III.  
In Sec.~IV, we discuss whether we can distinguish between the GMGHS black hole and 
the RN black hole by neutral particle collisions taking black hole shadow observations into account.
We conclude our results in Sec.~V.
We use units in which the speed of light and Newton's gravitational constant are unity.

\section{Asymptotically flat, static, and spherical symmetric spacetime}
We consider a general asymptotically flat, static, and spherical symmetric spacetime with a positive Arnowitt-Deser-Misner mass $M>0$, 
described by a line element 
\begin{equation}
\mathrm{d}s^2
=- f\left( r \right) \mathrm{d}t^2 +\frac{\mathrm{d}r^2}{f(r)} +C(r) \left( \mathrm{d}\vartheta^2 +\sin^2\vartheta \mathrm{d}\varphi^2 \right)\,.
\label{eq:line}
\end{equation}
The functions $f(r)$ and $C(r)$ satisfy
\begin{eqnarray}\label{eq:asymptotically-flat1}
\lim_{r \rightarrow \infty} f(r) = 1 - \frac{2M}{r} + O(r^{-2}) 
\end{eqnarray}
and 
\begin{eqnarray}\label{eq:asymptotically-flat2}
\lim_{r \rightarrow \infty} C(r) = r^2 +O(r)\,.
\end{eqnarray} 
Hereafter, we assume that $f(r)>0$ and $C(r)>0$ for $r>r_\mathrm{H}$ and 
$f(r_\mathrm{H})=0$, 
$C(r_\mathrm{H})>0$, 
and $f^\prime(r_\mathrm{H})>0$, where $r=r_\mathrm{H}$ is an event horizon
and the prime represents the differentiation with respect to $r$.

\subsection{Motion of a neutral particle}
We consider a particle motion with a mass $m$ and four-momentum 
\begin{eqnarray}
p^\mu=\dot{x}^\mu,
\end{eqnarray}
where the dot is the differentiation with respect to a parameter along the geodesic of the particle.
For simplicity, we assume that the particle moves on an equatorial plane $\vartheta=\pi/2$. 
We get the $t$ and $\varphi$ components of the four-momentum as  
\begin{eqnarray}\label{eq:pt}
p^t=\dot{t}=\frac{E}{f(r)}
\end{eqnarray}
and 
\begin{eqnarray}\label{eq:pvarphi}
p^\varphi=\dot{\varphi}=\frac{L}{C(r)}\,.
\end{eqnarray}
Here, $E\equiv-p_\mu t^\mu = f(r) p^t$ and $L\equiv p_\mu \varphi^\mu = C(r) p^\varphi$ are conserved energy and angular momentum of the particle where
\begin{equation}
t^\mu \partial_\mu = \partial_t
\end{equation}
and
\begin{equation}
\varphi^\mu \partial_\mu=\partial_\varphi 
\end{equation}
are time-translational and axial Killing vectors  
associated with the stationarity and axisymmetry of the spacetime, respectively.
From 
\begin{equation}\label{eq:pp}
-m^2=p^{\mu} p_{\mu}\,,
\end{equation}
the $r$ component of the four-momenta is obtained as
\begin{eqnarray}\label{eq:pr}
p^r=\dot{r}=\gamma \sqrt{-V(r)}\,,
\end{eqnarray}
where $\gamma$ is $1$ ($-1$) if the particle is outgoing (ingoing), and $V(r)$ is the effective potential of the radial motion of the particle defined by 
\begin{equation}\label{eq:defV}
V(r)=f(r)m^2-E^2+\frac{f(r)}{C(r)}L^2\,.
\end{equation}
From Eqs.~(\ref{eq:asymptotically-flat1}) and (\ref{eq:asymptotically-flat2}),
we get $V(\infty)=m^2-E^2$. Thus, particles with $m \leq E$ can be at a spatial infinity.
The motion of the particle on the equatorial plane $\vartheta=\pi/2$ is determined by Eqs.~(\ref{eq:pt}), (\ref{eq:pvarphi}), and (\ref{eq:pr}). 
The particle can be in domains where the effective potential is nonpositive and the particle is reflected at the zero points of the effective potential.

We consider a condition so that a particle with a marginal conserved energy $E=m$ reaches the event horizon at $r=r_\mathrm{H}$ from a spatial infinity $r \rightarrow \infty$.
The behavior of the effective potential on the event horizon and the spacial infinity is determined regardless the values of the angular momentum $L$ and the mass $m$ due to 
\begin{eqnarray}
&&V(r_\mathrm{H})=-m^2<0\,, \\
&&V(\infty) = \lim_{r \rightarrow \infty} -\frac{2M m^2}{r} <0\,, \\
&&V^\prime (r_\mathrm{H})= \left( f^\prime (r_\mathrm{H})+ \frac{f^\prime (r_\mathrm{H})}{C(r_\mathrm{H})} \right) m^2>0\,, \\
&&V^\prime (\infty) = \lim_{r \rightarrow \infty} \frac{2M m^2}{r^2} >0.
\end{eqnarray} 
Let $r_{\mathrm{c}}$ is a radius at which $V(r)$ takes its maximum in a domain $r_{\mathrm{H}} \leq r < \infty$.
We call $r_{\mathrm{c}}$ the critical radius. 
The particle reaches to the event horizon from the spacial infinity if and only if an inequality    
\begin{equation}
V(r_{\mathrm{c}})<0\,,
\end{equation}
holds.
On using Eq.~(\ref{eq:defV}), the inequality can be 
interpreted as a condition for the angular momentum, i.e., $\left| L \right| < L_{\mathrm{c}}$, 
where a critical angular momentum $L_{\mathrm{c}}$ is  defined as 
a positive angular momentum satisfying
\begin{equation}
V(r_{\mathrm{c}})=0\,.
\label{defLc}
\end{equation}
Under our assumptions, 
\begin{equation}
V'(r_{\mathrm{c}})=0\,,\label{defrc}
\end{equation}
must be satisfied at the critical radius.

\subsection{Neutral particle collision near the black hole}
We consider the collision of two neutral particles, which we name particles $1$ and $2$, moving on the equatorial plane $\theta=\pi/2$.
We assume that $C(r)$ is finite at a collision point $r$.
The square of the center-of-mass energy of particles $1$ and $2$ at the collision point $r$ is defined by 
\begin{equation}
E_\mathrm{CM}^2(r) \equiv -\left( p^{\mu}_1+ p^\mu_2 \right) \left( p_{1 \mu}+ p_{2 \mu} \right),
\end{equation}
where $p^{\mu}_1$ and $p^\mu_2$ are the four-momenta of the particles $1$ and $2$, respectively.
Hereinafter, physical quantities with the subscripts $1$ and $2$ denote the ones for particles $1$ and $2$, respectively. 
Note that the center-of-mass energy $E_\mathrm{CM}(r)$ is a scalar and that it is a function of the collision point $r$. 
From Eqs.~(\ref{eq:pt}), (\ref{eq:pvarphi}), (\ref{eq:pp}), and (\ref{eq:pr}),
we obtain 
\begin{eqnarray}\label{eq:Ecm0}
E_\mathrm{CM}^2(r) 
&=& m_1^2+m_2^2 -\frac{2 L_1 L_2}{C(r)} \nonumber\\
&&+\frac{2E_1 E_2- 2\gamma_1 \gamma_2 \sqrt{(-V_1(r))(-V_2(r))}}{f(r)}\,,  \nonumber\\ 
\end{eqnarray}
where the effective potential $V_i(r)$ for the particle~$i=1$ and $2$ is defined by 
\begin{equation}
V_i(r) \equiv f(r)m_i^2- E_i^2+\frac{f(r)}{C(r)}L_i^2\,. 
\end{equation}

For a rear-end collision $\gamma_1=\gamma_2=-1$,  
the numerator and denominator of the fourth term of Eq.~(\ref{eq:Ecm0}) vanish in a near-horizon limit $r \rightarrow r_\mathrm{H}+0$.
From l'Hospital's rule, this term becomes 
\begin{eqnarray}
\hspace{-.5cm}
&&\lim_{r\rightarrow r_\mathrm{H}+0 } \frac{2E_1 E_2- 2\sqrt{(-V_1(r))(-V_2(r))}}{f(r)} \nonumber\\
\hspace{-.5cm}
&=& \lim_{r\rightarrow r_\mathrm{H}+0 } \frac{ \left( 2E_1 E_2- 2 \sqrt{(-V_1(r))(-V_2(r))} \right)^\prime}{f^\prime(r)} \nonumber\\
\hspace{-.5cm}
&=& \left( m_1^2 +\frac{L_1^2}{C(r_\mathrm{H})} \right) \frac{E_2}{E_1}+\left( m_2^2 +\frac{L_2^2}{C(r_\mathrm{H})} \right) \frac{E_1}{E_2}\,,
\end{eqnarray}
and we obtain the square of the center-of-mass energy in the near-horizon limit as
\begin{eqnarray}\label{eq:Ecm0a}
\hspace{-.5cm}
E_\mathrm{CM}^2(r_\mathrm{H}) 
&=&m_1^2 \left( 1+\frac{E_2}{E_1} \right) +m_2^2 \left( 1+\frac{E_1}{E_2} \right)  \nonumber\\
&&+\frac{1}{C(r_\mathrm{H})} \left( L_1 \sqrt{\frac{E_2}{E_1}}- L_2 \sqrt{\frac{E_1}{E_2}} \right)^2.
\end{eqnarray}
For the rear-end collision of the marginal particles with the equal masses $m$ defined by $m\equiv E_1=E_2=m_1=m_2$,
the center-of-mass energy in the near-horizon limit is given by 
\begin{eqnarray}\label{eq:Ecm0b}
E_\mathrm{CM}(r_\mathrm{H}) 
=2m \sqrt{1+\frac{\left( l_1-l_2 \right)^2}{4C(r_\mathrm{H})}}\,, 
\end{eqnarray}
where $l_i$ is the specific angular momentum defined by $l_i \equiv L_i/m_i$ for particle $i=1$ and $2$. 
If the particles have their critical angular momenta with different signs, i.e., $L_1=\pm L_\mathrm{c}$ and $L_2=\mp L_\mathrm{c}$,
it reduces to 
\begin{equation}\label{eq:Ecm}
E_\mathrm{CM}(r_\mathrm{H}) = 2m \sqrt{1 +\frac{ l_\mathrm{c}^2 }{C(r_\mathrm{H})}}\,,
\end{equation}
where $l_{\mathrm{c}}$ is the critical angular momentum per unit rest mass 
of the particle defined by $l_{\mathrm{c}} \equiv L_{\mathrm{c}}/m$. 

If we set $m=m_1=m_2$, $E_1=E_2$, $L_1=L_2$, and $\gamma_1=\gamma_2$, the particles $1$ and $2$ are kinematically the same and their relative velocity vanishes. Then, we get the center-of-mass energy at a position $r$ as  
\begin{equation}\label{eq:Ecmsame}
E_\mathrm{CM}(r) = 2m\,.
\end{equation}
This trivial result is useful to check calculations.

\subsection{Motion of a photon}
If we set the mass of a particle $m=0$ and read a four momentum $p^\mu$ of the particle 
as the four wave number $k^\mu=\dot{x}^\mu$ of a photon, Eq.~(\ref{eq:pp}) reduces to 
$k^\mu k_\mu =0$. This relation gives the equation motion of the photon as 
\begin{equation}\label{eq:ray1}
-f(r)\dot{t}^2+ \frac{\dot{r}^2}{f(r)}+C(r) \dot{\varphi}^2=0\,.
\end{equation}
At the closest distance of the photon $r=r_0$, 
we get a relation 
\begin{equation}\label{eq:ray2}
f_0 \dot{t}_0^2=C_0 \dot{\varphi}_0^2\,,
\end{equation}
where functions with subscript $0$ denote ones at the closest distance,
since $\dot{r}$ vanishes there.
We define the impact parameter $b$ of the photon 
by 
\begin{equation}
b\equiv \frac{L}{E}\,.
\end{equation}
Since the impact parameter is constant along the geodesic of the photon, 
it can be expressed as 
\begin{equation}\label{eq:ray3}
b=b_0=\frac{L_0}{E_0}=\frac{C_0 \dot{\varphi}_0}{f_0 \dot{t}_0}=\pm \sqrt{\frac{C_0}{f_0}}\,,
\end{equation}
where we should choose the sign $\pm$ as the same signs as $\dot{\varphi}$, $L$, and $b$. 

The equation motion of the photon~(\ref{eq:ray1}) can be rewritten in 
\begin{equation}\label{eq:ray4}
\dot{r}^2+V(r,b)=0\,,
\end{equation}
where $V(r,b)$ is the effective potential for the photon given by
\begin{equation}
V(r,b) = \left( -1+\frac{f(r)}{C(r)}b^2 \right) E^2\,.
\end{equation}
The first and second derivatives of the effective potential $V(r,b)$ with respect to $r$ are given by 
\begin{equation}
\frac{\partial V(r,b) }{\partial r}  = \left( \frac{f(r)}{C(r)} \right)^\prime b^2 E^2
\end{equation}
and
\begin{equation}
\frac{\partial^2 V(r,b)  }{\partial r^2} = \left( \frac{f(r)}{C(r)} \right)^{\prime \prime} b^2 E^2\,,
\end{equation}
respectively.
We assume that the spacetime has a photon sphere which is formed by unstable circular photon orbits at $r=r_\mathrm{ph}$,
where $r_\mathrm{ph}$ is obtained as the smallest positive solution of an equation 
\begin{equation}
\left( \frac{f(r_\mathrm{ph})}{C(r_\mathrm{ph})} \right)^\prime =0\,,
\end{equation}
satisfying the inequality 
\begin{equation}
\left( \frac{f(r_\mathrm{ph})}{C(r_\mathrm{ph})} \right)^{\prime \prime}<0\,,
\end{equation}
and it should be larger than the radius of the event horizon $r_\mathrm{H}<r_\mathrm{ph}$.
The unstable circular photon orbit has a critical impact parameter $b_\mathrm{ph}$ define by 
\begin{equation}\label{eq:bph}
b_\mathrm{ph}\equiv \pm \sqrt{\frac{C(r_\mathrm{ph})}{f(r_\mathrm{ph})}}\,.
\end{equation}
Under the above assumption, the effective potential $V(r,b)$ satisfies the following expressions   
\begin{equation}
V(r_\mathrm{ph},b_\mathrm{ph})=0\,,
\end{equation}
\begin{equation}
\frac{\partial V(r_\mathrm{ph},b_\mathrm{ph}) }{\partial r}=0\,,
\end{equation}
and
\begin{equation}
\frac{\partial^2 V(r_\mathrm{ph},b_\mathrm{ph})  }{\partial r^2}<0\,.
\end{equation}
We define the changing rate $\rho$ of the impact parameter $b_\mathrm{ph}$ for the photon sphere by  
\begin{eqnarray}\label{eq:rhoDef}
\rho 
&\equiv& \frac{b_\mathrm{ph}}{b_{\mathrm{phSch}}}\,,
\end{eqnarray}
where $b_{\mathrm{phSch}}$ is the critical impact parameter in a Schwarzschild spacetime.

\section{Black hole spacetimes}

\subsection{Schwarzschild black hole}
The line element of the Schwarzschild black hole spacetime, which is known as the static, spherical symmetric vacuum solution of Einstein equations, is given by
\begin{eqnarray}
\mathrm{d}s^2
=- \left( 1-\frac{2M}{r} \right) \mathrm{d}t^2 +\left( 1-\frac{2M}{r} \right)^{-1}\mathrm{d}r^2 \nonumber\\ 
+ r^2 \left( \mathrm{d}\vartheta^2 +\sin^2\vartheta \mathrm{d}\varphi^2 \right)\,, 
\end{eqnarray}
which can be obtained from Eq.~(\ref{eq:line}) by setting $f(r)=1-2M/r$ and $C(r)=r^2$, where $M$ is the mass of the black hole.
The event horizon is at $r=r_\mathrm{H}=2M$. The critical radius of the marginal particle with $E=m$ 
is expressed as 
$r_\mathrm{c}=(l^2+\sqrt{l^4-12 l^2 M^2})/(2 M)$ 
from Eq.~(\ref{defrc}). Then, from Eq.~(\ref{defLc}), the
specific conserved angular momentum $l_\mathrm{c}$ per unit mass of the marginal particle is given by 
$l_\mathrm{c} = 4M$.

The center-of-mass energy (\ref{eq:Ecm0b}) in the near-horizon limit 
for the rear-end collision of the marginal particles with the equal masses 
is obtained as  
\begin{eqnarray}\label{eq:Ecm0bS}
E_\mathrm{CM}(r_\mathrm{H}) 
&=&2m \sqrt{1+\frac{\left( l_1-l_2 \right)^2}{16M^2}} \nonumber\\
&=&\sqrt{2}m \sqrt{\frac{16M^2+\left( l_1-l_2 \right)^2}{8M^2}}\,.
\end{eqnarray}
If we also assume that the particles have the critical conserved angular momenta with the different signs, 
the center-of-mass energy~(\ref{eq:Ecm}) reduces to    
\begin{equation}\label{eq:Ecm0bS2}
E_\mathrm{CM}(r_\mathrm{H}) = 2 \sqrt{5} m\,.
\end{equation}
Equation~(\ref{eq:Ecm0bS2}) is consistent with a result by Baushev~\cite{Baushev:2008yz}.

Pradhan got the center-of-mass energy in the Schwarzschild spacetime as Eq.~(38) in Ref.~\cite{Pradhan:2013hxa} which contradicts our result~(\ref{eq:Ecm0bS}).
In our notation, Eq.~(38) in Ref.~\cite{Pradhan:2013hxa} is expressed by 
\begin{eqnarray}\label{eq:Ecmwrong2}
E_\mathrm{CM}(r_\mathrm{H}) 
=\sqrt{2}m \sqrt{\frac{16M^2 +  \left( l_1-l_2 \right)^2}{ 4M^2 }}\,.
\end{eqnarray}
If we assume the particles have the critical conserved angular momenta with the different signs, i.e., 
$l_1=\pm l_c= \pm 4M$ and $l_2 = \mp l_c = \mp 4M$,
the center-of-mass energy~(\ref{eq:Ecmwrong2}) reduces to 
\begin{equation}
E_\mathrm{CM}(r_\mathrm{H}) = 2 \sqrt{10} m
\end{equation}
which contradicts both Eq.~(\ref{eq:Ecm0bS2}) in this paper and the well-known result in Ref.~\cite{Baushev:2008yz}.
Therefore, Eq.~(38) in Ref.~\cite{Pradhan:2013hxa} should be read as Eq.~(\ref{eq:Ecm0b}) in this paper.

The Schwarzschild spacetime has the photon sphere at
\begin{equation}
r=r_\mathrm{ph}=3M
\end{equation}
and 
the critical impact parameter for the photon sphere is obtained as 
\begin{equation}
b_{\mathrm{phSch}} \equiv b_\mathrm{ph} = 3 \sqrt{3} M\,.
\end{equation}

\subsection{GMGHS black hole}
\begin{figure}[htbp]
\begin{center}
\includegraphics[width=\linewidth]{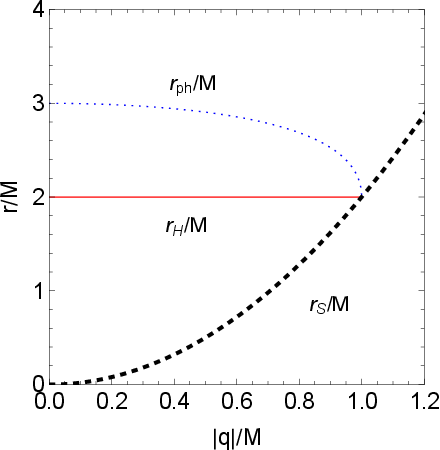}
\end{center}
\caption{
\label{fig:radii} 
The specific radii of the photon sphere $r_\mathrm{ph}/M$, the event horizon $r_\mathrm{H}/M$, the singularity $r_\mathrm{S}/M$  
against the specific charge $\left| q \right|/M$ are shown. 
Thin-dotted~(blue), thin-solid~(red), and thick-dashed~(black) curves denote $r_\mathrm{ph}/M$, $r_\mathrm{H}/M$, and $r_\mathrm{S}/M$, respectively.}
\end{figure}

We consider an Einstein-Maxwell-dilaton theory with an action~\cite{Gibbons:1982ih,Gibbons:1987ps,Garfinkle:1990qj} 
\begin{equation}
S=\int d^4 x \sqrt{-g} \left\{ -R +2 (\nabla \phi)^2 + e^{-2 \phi} F^2 \right\},
\end{equation}
where $R$ is a Ricci scalar, $\phi$ is a dilaton field, and $(\nabla \phi)^2=\nabla_\mu \phi \nabla^\mu \phi$, $F^2=F^{\mu \nu} F_{\mu \nu}$, and $F_{\mu \nu}=\nabla_\mu A_\nu-\nabla_\nu A_\mu$ with the vector potential $A_{\mu}$.
From the action, the field equations with respect to $A_\mu$, $\phi$, and $g_{\mu \nu}$, are obtained as
\begin{equation}
\nabla_\mu (e^{-2 \phi} F^{\mu \nu} ) =0\,,
\end{equation}
\begin{equation}
\nabla^2 \phi + \frac{1}{2} e^{-2 \phi} F^{2} =0\,,
\end{equation}
and
\begin{equation}
R_{\mu \nu}= 2\nabla_\mu \phi \nabla_\nu \phi + 2e^{-2 \phi} F_{\mu \rho} F_\nu^\rho - \frac{1}{2} g_{\mu \nu} e^{-2 \phi} F^2\,,
\end{equation}
respectively.
By a straightforward calculation, an asymptotically flat, static and spherically symmetric black hole solution with 
a purely magnetic Maxwell field
\begin{equation}
F=\tilde{Q} \sin \theta d\theta \wedge d \varphi\,, 
\end{equation}
where $\tilde{Q}$ is a magnetic charge\footnote{This electromagnetic tensor corresponds to the choice of the vector potential 
$A_{\mu}=(0,0,0,-\tilde{Q}\cos\theta)$.},
called a GMGHS black hole solution~\cite{Gibbons:1987ps,Garfinkle:1990qj},
is obtained in the coordinates $x^\mu=(t, r, \vartheta, \varphi)$ as
\begin{eqnarray}
\mathrm{d}s^2
&=& -\left(1-\frac{2M}{r} \right)\mathrm{d}t^2 + \left( 1-\frac{2M}{r} \right)^{-1}\mathrm{d}r^2 \nonumber\\
&&+ r \left(r -\frac{2q^2}{M} \right) \left( \mathrm{d}\vartheta^2 +\sin^2 \vartheta \mathrm{d}\varphi^2 \right)\,,
\end{eqnarray}
\begin{equation}
e^{-2\phi(r)}=\frac{e^{-2\phi_0}}{r} \left(r-\frac{2q^2}{M} \right)\,,
\end{equation}
and
\begin{equation}
F= \sqrt{2} q e^{\phi_0} \sin \theta d\theta \wedge d \varphi\,,
\end{equation}
where $q=\tilde{Q} e^{-\phi_0}/\sqrt{2}$
and we have set a constant $\phi_0$ to realize $\phi_0=\phi(r \rightarrow \infty)$.
\footnote{
In Ref.~\cite{EventHorizonTelescope:2021dqv}, 
a charge was introduced $\bar{q}\equiv \tilde{Q} e^{-\phi_0}$ in our notation and  $q=\bar{q}/\sqrt{2}$ was called a normalized physical charge.  
The normalized charge $q$ in our notation was used in Fig.~$2$~(left) in Ref.~\cite{EventHorizonTelescope:2021dqv} and Fig.~$18$~(left) in Ref.~\cite{EventHorizonTelescope:2022xqj} 
and the charge $\bar{q}$ was used in Fig.~$28$ in Ref.~\cite{Vagnozzi:2022moj}.
}
The Ricci scalar $R$ is given by
\begin{equation}
R=\frac{2q^4}{M^2 r^3} \frac{r-2M}{ \left( r-{2q^2}/{M} \right)^2 }
\end{equation}
which diverges at $r=2q^2/M$. 
The spacetime has an event horizon at $r=r_\mathrm{H}=2M$ for $\left| q \right|<M$,
naked singularity at $r=r_\mathrm{S}=2q^2/M$ for $\left| q \right|>M$, 
and the horizon and the curvature singularity coincides at $r=2M$ for $\left| q \right|=M$.
The last case characterized by $\left| q \right|=M$ corresponds to the extreme charged 
black hole for which the Ricci scalar reduces to 
\begin{equation}
R=\frac{2M^2}{r^3 \left( r-2M \right) }\,.
\end{equation}
The coincidence of the curvature singularity and the event horizon is the pronounced feature of the GMGHS black hole
and it can be used to distinguish it from other black hole solutions.  
The spacetime has a photon sphere at
\begin{equation}
r=r_\mathrm{ph}=\frac{3 M^2+q^2+\sqrt{9 M^4-10 M^2 q^2+q^4}}{2 M}
\end{equation}
for $\left| q \right|< M$.
The radii $r_\mathrm{ph}$, $r_\mathrm{H}$, and $r_\mathrm{S}$ are shown in Fig.~\ref{fig:radii}. 

\begin{figure*}[htbp]
\begin{center}
\includegraphics[width=60mm]{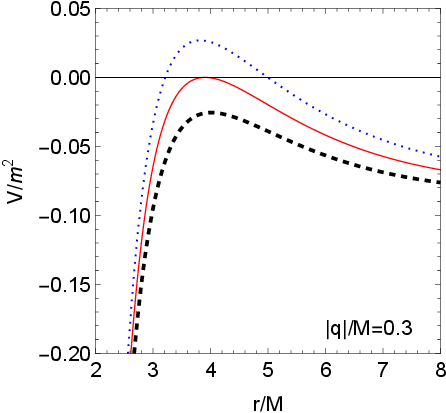}
\includegraphics[width=58mm]{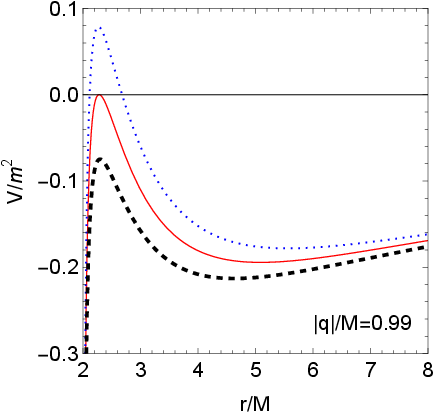}
\includegraphics[width=58mm]{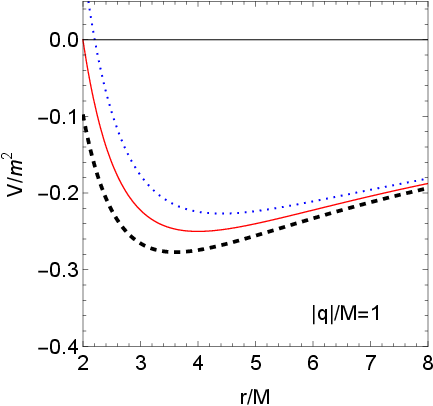}
\end{center}
\caption{
\label{fig:V13} 
The dimensionless effective potentials $V/m^2$ with $E=m$ are shown as a function of a dimensionless radius $r/M$ in the GMGHS spacetime with $\left| q \right|=0.3M$, $0.99M$, and $M$ in left, middle, and right panels are shown, respectively. 
Thin-dotted~(blue), thin-solid~(red), and thick-dashed~(black) curves denote $V/m^2$ with $L=L_\mathrm{c}+0.1Mm$, $L_\mathrm{c}$, and $L_\mathrm{c}-0.1Mm$. 
Every panel is plotted in a domain for $r_\mathrm{H}/M \leq r/M \leq 8$. 
}
\end{figure*}
In the GMGHS black hole spacetime, 
the effective potential $V(r)$ for the particles are shown in Fig.~\ref{fig:V13} and  
$C(r_\mathrm{H})$ is given by
\begin{equation}
C(r_\mathrm{H})=4\left( M^2-q^2 \right)\,. 
\end{equation}
The critical specific conserved angular momentum per unit mass $l_\mathrm{c}$ 
for the marginal particle with $E=m$ is obtained as 
\begin{equation}
l_\mathrm{c} = 2 \left(M + \sqrt{M^2-q^2} \right)\,.
\end{equation}

\begin{figure}[htbp]
\begin{center}
\includegraphics[width=80mm]{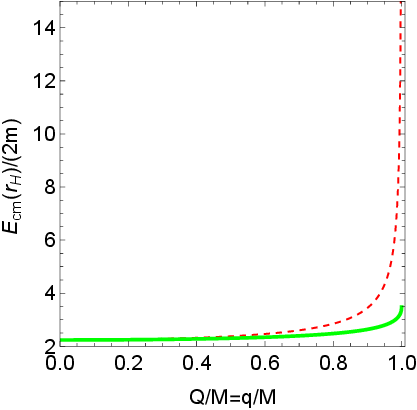}
\end{center}
\caption{
\label{fig:EcmQ} 
$E_\mathrm{CM}(r_\mathrm{H})/(2m)$ in GMGHS and RN spacetimes are shown.
Thin-dashed (red) and thick-solid (green) curves denote Eqs.~(\ref{eq:Ecm1}) and (\ref{eq:Ecm2}), respectively. 
We have set $Q/M=q/M$ to compare them.
}
\end{figure}

For the rear-end collision of the marginal particles with the equal masses,
the center-of-mass energy~(\ref{eq:Ecm0b}) in the near-horizon limit 
can be expressed~\footnote{
The alternative expressions of Eq.~(\ref{eq:Ecm0bGM}) 
\begin{eqnarray}
E_\mathrm{CM}(r_\mathrm{H}) 
&=&\sqrt{2}m \sqrt{\frac{8M \left( 2M-2q^2/M \right) +  \left( l_1-l_2 \right)^2}{4M \left( 2M-2q^2/M \right) }} \nonumber
\end{eqnarray}
would be useful to compare it and expressions in Refs.~\cite{Pradhan:2013hxa,Fernando:2013qba} discussed later.}

\begin{eqnarray}\label{eq:Ecm0bGM}
E_\mathrm{CM}(r_\mathrm{H}) 
=2m \sqrt{1+\frac{\left( l_1-l_2 \right)^2}{16 \left( M^2-q^2 \right) }} \,. 
\end{eqnarray}
The center-of-mass energy~(\ref{eq:Ecm}) for the marginal particles having the critical angular momenta with different signs is given by 
\begin{eqnarray}\label{eq:Ecm1}
\frac{E_\mathrm{CM}(r_\mathrm{H})}{2m} 
=\sqrt{1+\frac{\left(\sqrt{M^2-q^2}+M \right)^2}{M^2-q^2}}
\end{eqnarray}
and it is shown in Fig.~\ref{fig:EcmQ}.  

Pradhan obtained the center-of-mass energy for the rear-end collision of particles with $m=m_1=m_2=E_1=E_2$ in the near-horizon limit as Eqs.~(37), (39), and (40) in Ref.~\cite{Pradhan:2013hxa} which contradict our result~(\ref{eq:Ecm0b}).
In our notation, Eqs.~(37), (39), and (40) in Ref.~\cite{Pradhan:2013hxa} can be expressed by 
\begin{eqnarray}\label{eq:Ecmwrong}
E_\mathrm{CM}(r_\mathrm{H}) 
&=&\sqrt{2}m \sqrt{\frac{8M\left( 2M-{2q^2}/{M} \right)+  \left( l_1-l_2 \right)^2}{ 2M \left( 2M -{2q^2}/{M} \right)}}\,.\nonumber\\
\end{eqnarray}
When the particles have the same angular momenta $l_1=l_2$, Eqs.~(37), (39), and (40) in Ref.~\cite{Pradhan:2013hxa} or Eq.~(\ref{eq:Ecmwrong}) yields $E_\mathrm{CM}(r_\mathrm{H}) = 2\sqrt{2}m$ and it does not recover Eq.~(\ref{eq:Ecmsame}) or $E_\mathrm{CM}(r_\mathrm{H}) = 2m$.
Thus, Eqs.~(37), (39), and (40) in Ref.~\cite{Pradhan:2013hxa} should be read as Eq.~(\ref{eq:Ecm0b}) in this paper.
We also note that the square of the center-of-mass energy $E_\mathrm{CM}^2(r)$ for the rear-end collision of the particles with $m_1=m_2=E_1=E_2$ was obtained as 
Eq.~(32) in Ref.~\cite{Fernando:2013qba}.
Fernando called the square of the center-of-mass energy the center-of-mass energy.
Thus, if we modified the definition of the center-of-mass energy in Ref.~\cite{Fernando:2013qba}, 
we can confirm that Eq.~(\ref{eq:Ecm0b}) in this paper is consistent with the modified Eq.~(32) in Ref.~\cite{Fernando:2013qba}.

In the GMGHS spacetime, 
the impact parameter for the photon sphere is given by $b=b_\mathrm{ph}$, where 
\begin{eqnarray}\label{eq:bph1}
\hspace{-1cm}
b_\mathrm{ph}
&=&\frac{1}{\sqrt{2}M}\left[
27M^4-18M^2q^2-q^4\right.\nonumber\\
&&\left.
+(9M^2-q^2)\sqrt{9M^4-10M^2q^2+q^4}
\right]^{1/2}\,, 
\end{eqnarray}
and the changing rate $\rho$ of the impact parameter $b_\mathrm{ph}$ is given by  
\begin{eqnarray}\label{eq:rho1}
\hspace{-1cm}
\rho 
&=&\frac{1}{3 \sqrt{6} M^2}
\left[
27M^4-18M^2q^2-q^4\right.\nonumber\\
&&\left.
+(9M^2-q^2)\sqrt{9M^4-10M^2q^2+q^4}
\right]^{1/2}\,.
\end{eqnarray}

\subsection{Reissner-Nordstr\"{o}m black hole}
The metric of a Reissner-Nordstr\"{o}m spacetime is given by 
\begin{eqnarray}
\mathrm{d}s^2
&=&- \left( 1- \frac{2M}{r} + \frac{Q^2}{r^2} \right) \mathrm{d}t^2
+\left({1- \frac{2M}{r} + \frac{Q^2}{r^2}}\right)^{-1}{\mathrm{d}r^2} \nonumber\\
&&+r^2 ( \mathrm{d}\vartheta^2 +\sin^2\vartheta \mathrm{d}\varphi^2)\,,
\end{eqnarray}
where $Q$ is the electrical charge of a central object. 
This line element corresponds to $f(r)= 1- {2M}/{r} + {Q^2}/{r^2}$ and $C(r)=r^2$ in Eq.~(\ref{eq:line}).
The spacetime has a curvature singularity at $r=r_\mathrm{S}=0$.
It has an event horizon at 
\begin{equation}\label{eq:rH}
r=r_\mathrm{H}\equiv M +\sqrt{M^2-Q^2}\,,
\end{equation}
for $\left| Q \right| \geq M$ 
and naked singularity for $\left| Q \right| > M$. 

\begin{figure*}[htbp]
\begin{center}
\includegraphics[width=59mm]{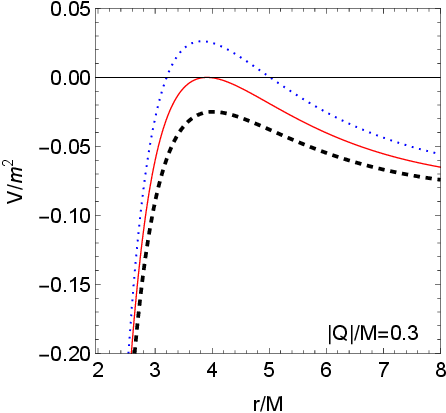}
\includegraphics[width=59mm]{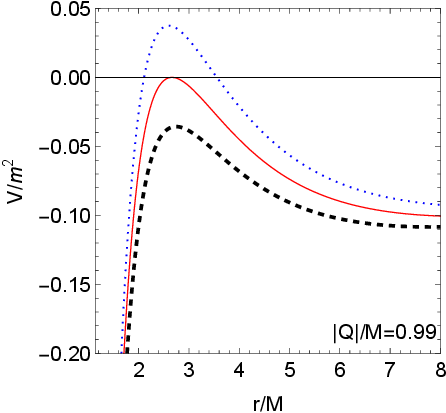}
\includegraphics[width=59mm]{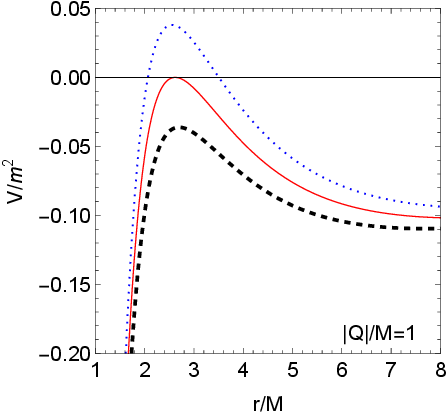}
\end{center}
\caption{
\label{fig:V46} 
The dimensionless effective potentials $V/m^2$ with $E=m$ are shown as a function of a dimensionless radius $r/M$ in the Reissner-Nordstr\"{o}m spacetime with $\left| Q \right|=0.3M$, $0.99M$, and $M$ in left, middle, and right panels are shown, respectively. 
Thin-dotted~(blue), thin-solid~(red), and thick-dashed~(black) curves denote $V/m^2$ with $L=L_\mathrm{c}+0.1Mm$, $L_\mathrm{c}$, and $L_\mathrm{c}-0.1Mm$. 
Every panel is plotted in a domain for $r_\mathrm{H}/M \leq r/M \leq 8$. 
}
\end{figure*}

In the Reissner-Nordstr\"{o}m spacetime, 
the effective potential $V(r)$ for neutral particles is plotted in Fig.~\ref{fig:V46}. 
The expression of critical conserved angular momentum per unit mass 
\begin{equation}
l_{\mathrm{c}} = l_{\mathrm{cRN}}
\end{equation}
is shown in appendix~A.

The center-of-mass energy~(\ref{eq:Ecm0b}) in the near-horizon limit for the rear-end collision of the marginal particles with the equal masses is given by 
\begin{eqnarray}
E_\mathrm{CM}(r_\mathrm{H}) 
&=&2m \sqrt{1+\frac{\left( l_1-l_2 \right)^2}{4r_\mathrm{H}^2}} 
\end{eqnarray}
and it is consistent with Eq.~(55) in Ref.~\cite{Pradhan:2013hxa}.
The center-of-mass energy~(\ref{eq:Ecm}) for the marginal particles with the equal masses and the different signed critical angular momenta $l_1=\pm l_\mathrm{c}$ and $l_2=\mp l_\mathrm{c}$
is given by
\begin{eqnarray}\label{eq:Ecm2}
\frac{E_\mathrm{CM}(r_\mathrm{H})}{2m} 
=\sqrt{1+\frac{l_{\mathrm{cRN}}^2}{r_\mathrm{H}^2}}
\end{eqnarray}
and it is plotted in Fig.~\ref{fig:EcmQ}.  

The RN spacetime has 
the photon sphere at $r=r_\mathrm{ph}$, where $r_\mathrm{ph}$ is given by
\begin{equation}
r_\mathrm{ph}=\frac{3 M+\sqrt{9 M^2-8Q^2}}{2}
\end{equation}
for $\left| Q \right|< 3M/(2\sqrt{2})$.
The impact parameter for the photon sphere (\ref{eq:bph}) reduces to 
\begin{eqnarray}\label{eq:bph2}
b_\mathrm{ph}
=\frac{3 M \left(\sqrt{9 M^2-8 Q^2}+3 M\right)-4 Q^2}{\sqrt{2 M \left(\sqrt{9 M^2-8 Q^2}+3 M\right)-4 Q^2}}\,.
\end{eqnarray}
In the RN spacetime, the changing rate of the impact parameter $b_\mathrm{ph}$ for the photon sphere is obtained by   
\begin{equation}\label{eq:rho2}
\rho =\frac{3 M \left(\sqrt{9 M^2-8 Q^2}+3 M\right)-4 Q^2}{3 \sqrt{3} M \sqrt{2 M \left(\sqrt{9 M^2-8 Q^2}+3 M\right)-4 Q^2}}\,.
\end{equation}

\section{Discussion and Conclusion}
The center-of-mass energy is used to indicate the performance of a particle accelerator. For an example, the Large Hadron Collider was designed to reach 14.0 TeV of the center-of-mass energy, which is $1.5 \times 10^4$ times the rest mass of a proton~\cite{Harada:2014vka}. 
Kinematically, the center-of-mass energy means the upperbound of rest mass of the creation after the particle collision. This implies that the BSW collision can potentially open a window of high energy physics which we cannot reach by terrestrial accelerators. 
However, the almost products of the BSW collision near the Kerr black hole will be fall into the black hole, and the rest of the ejecta, which can escape far away from the black hole, must be highly redshifted~\cite{Banados:2010kn}.
The particles with high energy or large mass produced by the BSW collision cannot escape from the extremal Kerr black hole~\cite{Bejger:2012yb,Harada:2012ap}. 
The observational effects of the extremal Kerr black hole have been investigated in Refs.~\cite{Banados:2010kn,Williams:2011uz,Cannoni:2012rv} showing that the effect on the flux is too small to be detected due to the strong redshift and the small escape fraction~\cite{McWilliams:2012nx,Zaslavskii:2013et,McWilliams:2013efa}. The possibility of the indirect observations has been discussed in Ref.~\cite{Gariel:2014ara}. 
The details of the indirect observational effects can depend on the class of particle collision with the high center-of-mass energy, spacetime, astrophysical situations, and so on. The indirect observational effects of the BSW collisions and other class of the particle collisions with arbitrary high center-of-mass energy can be an open question and further investigations are needed. 

It is widely believed that astrophysical black holes are neutralized by plasma around the black holes. 
On the other hand, it has been discussed that it would be natural for the black holes in some astrophysical situations to have a small amount of the electrical charge~\cite{Wald:1974np,Zajacek:2018ycb,Levin:2018mzg,King:2021jlb,Komissarov:2021vks,Nakao:2024wek}. 
To determine the electrical charge of the black holes from observations, we need to observe phenomena very close to the black hole, because the astrophysical environments around the black holes may screen the electrical charge of the black holes, and because the effects of electrical charge on the black hole metrics become weaker than gravity away from the black holes. 
We, however, emphasize that the black hole charge of the GMGHS metric is not necessarily neutralized by astrophysical environment around the black hole since it does not necessarily correspond to the electrical charge in electromagnetism. Recently, the EHT collaboration has put bounds on the black hole charges of M87*~\cite{EventHorizonTelescope:2021dqv} and Sgr~A*~\cite{EventHorizonTelescope:2022xqj} by assuming that the size of the observed ring is proportional to the changing rate of photon sphere. Their method for constraining the large class of black hole charges is very simple and it can be applied for both electrically charged and the other kinds of charged black holes including the GMGHS black holes.

Let us consider how do we compare the center-of-mass energy for the collision of neutral particles in a spacetime with a charge 
against that in another spacetime with a different kind of charge?
We suggest three options to answer the question: 
\begin{description} 
 \item[(a)] We can compare the center-of-mass energy with the same values of different normalized charges as shown in Fig.~\ref{fig:EcmQ} and Appendix~B. 
We can apply this way for every spacetime easily but the validity of the comparison is uncertain.
 \item[(b)] If we have the relation of the different kind of charges, we can use it to compare the center-of-mass energy.  
For an example, from a black hole solution~\cite{Gibbons:1987ps,Gibbons:1985ac} including both the Reissner-Nordstr\"{o}m solution and the GMGHS solution, we can find the relation of the charges $q$ and $Q$. 
Thus, we can compare the center-of-mass energy by considering the spacetime including both spacetimes.  
However, this method is only applicable to very special spacetimes.
 \item[(c)] We suggest a new method to compare the center-of-mass energy in the spacetimes with the different 
kinds of charges by using shadow observations and mass observations.
The validity of this method is better than method (a) in the sense that we can adopt it 
regardless of the types of charges,
and we can apply it for every spacetime while the observational data exist only for Sgr~A* and M87* so far. 
\end{description} 
In the rest of this section, we concentrate on the method (c).

The distance and mass of Sgr~A* seem to be accurately estimated as
$D=8277 \pm 9 \pm 33$~pc and $M=(4.297 \pm 0.013) \times 10^6 M_\odot$ by Very Large Telescope Interferometer (VLTI) observations~\cite{GRAVITY:202101,GRAVITY:2021xju} 
and $D=7935 \pm 50 \pm 32$~pc and $M=(3.951 \pm 0.047) \times 10^6 M_\odot$ by Keck observations~\cite{Do:2019txf}.
The EHT collaboration reported a ring image of Sgr~A* with the diameter of $51.8 \pm 2.3~\mu$as (at $1 \sigma$ probability) from EHT observations at the wavelength of $1.3$~mm~\cite{EventHorizonTelescope:2022wkp}. 
They found $\delta=-0.08 \pm0.09$ from the VLTI observations and $\delta=-0.04^{+0.09}_{-0.10}$ from the Keck observations~\cite{EventHorizonTelescope:2022xqj},
where $\delta$ is defined by
\begin{equation}
\delta \equiv \frac{\theta_\mathrm{g}}{\theta}-1,
\end{equation}
and $\theta_\mathrm{g} \equiv M/D$ is a ratio of the mass $M$ to the distance $D$ estimated from the EHT observations by assuming the Kerr spacetime
while $\theta$ is the ratios $M/D$ estimated by using the VLTI observations or the Keck observations.  
They concluded the ring image of Sgr~A* is consistent with the Kerr black hole with the mass obtained from VLTI and Keck observations.  

\begin{figure*}[htbp]
\begin{center}
\includegraphics[width=59mm]{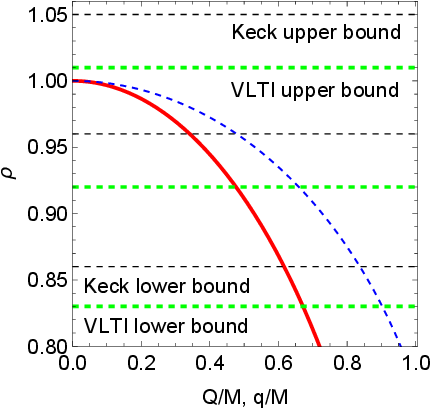}
\includegraphics[width=58mm]{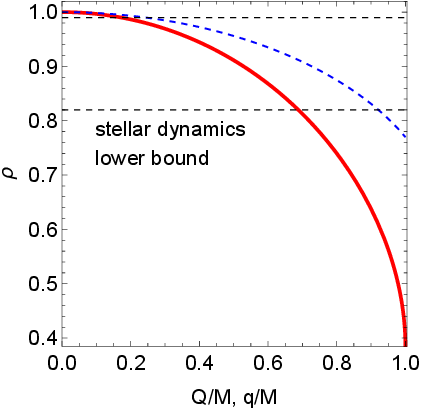}
\end{center}
\caption{
\label{fig:qr} 
The changing rates $\rho$ of photon spheres as the function of $q/M$ for the GMGHS spacetime denoted by the thin-solid (red) curve curve and $\rho$ as the function of $Q/M$ for the RN spacetime denoted by the thick-dashed (blue). The left and right panels show constraints from the observations of Sgr~A* and M87*, respectively. 
}
\end{figure*}

\begin{figure*}[htbp]
\begin{center}
\includegraphics[width=57mm]{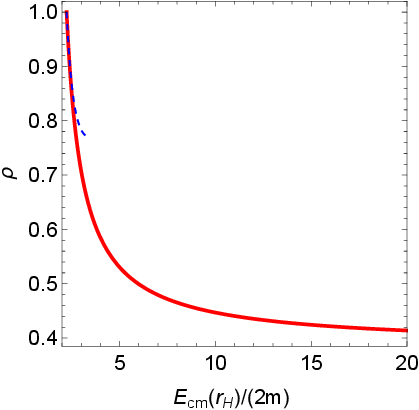}
\includegraphics[width=59mm]{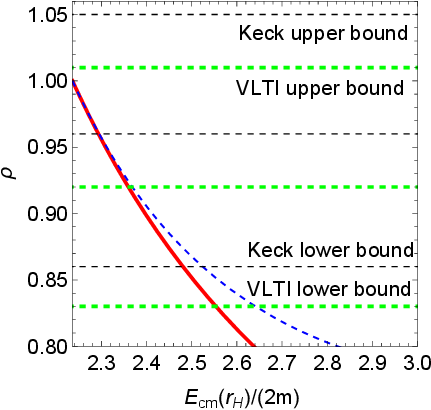}
\includegraphics[width=59mm]{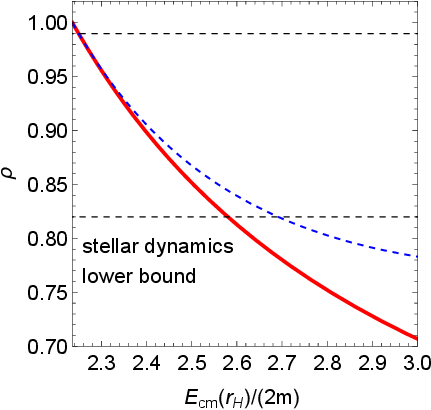}
\end{center}
\caption{
\label{fig:Ecmrho} 
A left panel shows the relation between the changing rates $\rho$ of photon spheres as the functions of $q /M$ and the center-of-mass energy $E_\mathrm{CM}(r_\mathrm{H})/(2m)$ for the GMGHS spacetime denoted by the thin-solid (red) curve and $\rho$ as the function of $Q/M$ for the RN spacetime denoted by the thick-dashed (blue).
Middle and left panels show the constraints on the center-of-mass energy $E_\mathrm{CM}(r_\mathrm{H})/(2m)$ from the observations of Sgr~A* and M87*, respectively. }
\end{figure*}

The EHT collaboration~\cite{EventHorizonTelescope:2022xqj} put constraints on the charges as $0 \leq \left| Q \right| /M \leq  0.84$ for RN black hole and as $0 \leq \left| q \right| /M \leq  0.62$ for the GMGHS black hole from the EHT observations and the Keck observations $\delta=-0.04^{+0.09}_{-0.10}$
under an assumption that the size of observed ring should be proportional to 
the size of the photon sphere which varies with the changing rate $\rho$ of the charge
as shown in Fig.~\ref{fig:qr}. 
Since the center-of-mass energy $E_\mathrm{CM}$ depends on the charge as Eq.~(\ref{eq:Ecm1}) in the GMGHS spacetime and as Eq.~(\ref{eq:Ecm2}) in the RN spacetime, the observational bound on the charge enables us to put constraint on $E_\mathrm{CM}$. 
We apply the bounds of the charges from the EHT observations~\cite{EventHorizonTelescope:2022xqj} and the mass observations by VLTI~\cite{GRAVITY:202101,GRAVITY:2021xju} and by Keck~\cite{Do:2019txf} to the center-of-mass energy. 
Due to the existence of the upper bound on the charges, the center-of-mass energy cannot extremely large. 
Consequently, we find that the maximum value of $E_\mathrm{CM}$ in GMGHS spacetime is allowed to be 
smaller than that in the RN spacetime as shown in Fig.~\ref{fig:Ecmrho}.
We conclude that the difference of center-of-mass energy in the two spacetimes is only within a few percent and 
that we cannot distinguish the difference for the Sgr~A*.

The mass of M87* was estimated as $M=(6.6 \pm 0.4) \times 10^9 M_\odot$ and $M=(3.5^{+0.9}_{-0.7}) \times 10^9 M_\odot$ (at $1 \sigma$ probability)  
from stellar-dynamics observations~\cite{Gebhardt:2011yw} and gas-dynamical measurements~\cite{Walsh:2013uua}, respectively, 
under the assumption of a distance of $D=17.9$ Mpc.
The EHT collaboration reported the ring image of M87* with a diameter of $42 \pm 3$~$\mu$as (at $1 \sigma$ probability) 
from observations at a wavelength of $1.3$ mm~\cite{EventHorizonTelescope:2019dse,EventHorizonTelescope:2019ggy} 
by assuming the Kerr spacetime~\cite{EventHorizonTelescope:2019ggy}.  
The EHT collaboration estimated the ratios $\theta$
and they found the values of $\delta=-0.01\pm 0.17$ and $0.78\pm 0.3$ (at $1 \sigma$ probability) 
from the stellar dynamics observations~\cite{Gebhardt:2011yw} and the gas dynamics observations~\cite{Walsh:2013uua}, respectively.
The EHT collaboration concluded that the observed ring image of M87* under the assumption of the Kerr black hole is consistent 
with the stellar dynamics observations~\cite{Gebhardt:2011yw} but not with gas-dynamical mass measurements~\cite{Walsh:2013uua}
\footnote{
The EHT collaboration adapted the distance $D=16.8 \pm 0.7$Mpc to obtain the mass $M=(6.5 \pm 0.7) \times 10^9  M_\odot$~\cite{EventHorizonTelescope:2019ggy}.
Recently, the mass of M87* has been revisited and estimated as $(5.37^{+0.37}_{-0.25} \pm 0.22) \times 10^9 M_\odot$ by \cite{Liepold2023}
and $(8.7 \pm 1.2 \pm 1.3 ) \times 10^9  M_\odot$ by \cite{Simon2024}
by assuming the distance of $D=16.8 \pm 0.7$Mpc.
}.
The EHT collaboration~\cite{EventHorizonTelescope:2021dqv} adapted not $\delta=-0.01\pm 0.17$ from EHT observations and the stellar dynamics observations 
but $\delta=0.00\pm 0.17$ to put constraints on the charges as $0 \leq \left| Q \right| /M \leq 0.90$ for the RN black hole 
and as $0 \leq \left| q \right| /M \leq 0.67$ for the GMGHS black hole
under the assumption that the size of observed ring should be proportional to that of the photon sphere.
Figure~\ref{fig:Ecmrho} implies that the difference of the center-of-mass energy in the spacetimes is less than $5$ percents and we cannot distinguish them. 
We cannot, however, make clear definite conclusions right now because our method requires precise and accurate mass and shadow measurements 
to determine the value of the center-of-mass energy while the stellar dynamics observations~\cite{Gebhardt:2011yw} 
and gas-dynamical mass measurements~\cite{Walsh:2013uua} of M87* are in significant disagreement.

We add a comment on certainty of constraints on the charges of M87*~\cite{EventHorizonTelescope:2021dqv} and Sgr~A*~\cite{EventHorizonTelescope:2022xqj}
since one may be curious whether (almost) extremal charged black holes can be permitted from current observations. 
The constraints on the charges of various black hole metric given by the EHT collaboration~\cite{EventHorizonTelescope:2021dqv,EventHorizonTelescope:2022xqj} 
depend only on the metric of the charged black holes because they used a simple assumption which does not involve ray-traced general-relativistic magnetohydrodynamic simulations for the charged black hole spacetimes. To obtain more reliable constraints on the charges from shadow observations, these simulations should be performed since the values of the charge must depend also on the matter field equations and gravitational theories. 
Future observations~\cite{Lupsasca:2024xhq,Johnson:2024ttr} may give important information to fix the values of charges since the charges also depend on astrophysical assumptions~\cite{Gralla:2020pra}.     
In a word, it is hard to exclude the (almost) extremal charged black holes with absolute certainty from the current observations and simulations, and we should check the assumptions carefully~\cite{Tsukamoto:2024gkz}.

\section*{Acknowledgements}
The authors are grateful to Kitaro Taniguchi and Yuta Suzuka for useful discussion 
and to an anonymous referee for useful comments.
N.T. thanks Yosuke Mizuno for his kind comment on a solution in Ref.~\cite{Mizuno:2018lxz}. 
R.K. is supported by the Grant-in-Aid for Scientific Research (C) of the JSPS No.~23K03421. 
\appendix
\section{$l_\mathrm{cRN}$}
We introduce specific conserved angular momentum and energy of the particle $l\equiv L/m$ and $e\equiv E/m$, respectively. We also introduce 
the dimensionless angular momentum of the particle $l_{*}\equiv l/M= L/(mM)$ and the dimensionless electric charge of the black hole $Q_{*}\equiv Q/M$.
The critical dimensionless angular momentum $l_\mathrm{*cRN}\equiv l_\mathrm{cRN}/M = L_\mathrm{cRN}/(mM)$ is obtained as a root of the equation
\begin{eqnarray}\label{eq:lcRN1}
&&l_{*}^8 \left(e^2-1\right) \left(Q_{*}^2-1\right) \nonumber\\
&&+l_{*}^6 \left[ e^4 \left(8 Q_{*}^4-36 Q_{*}^2+27\right)+e^2 \left(-12 Q_{*}^4+49 Q_{*}^2-36\right) \right. \nonumber\\
&&\left. +4 \left(Q_{*}^4-3 Q_{*}^2+2\right)\right] \nonumber\\
&&+l_{*}^4 \left[ 4 \left(3 e^2-8\right) Q_{*}^2+\left(12 e^4-31 e^2+22\right) Q_{*}^4 \right. \nonumber\\
&&\left.+2 \left(8 e^6-16 e^4+11 e^2-3\right) Q_{*}^6+16\right] \nonumber\\
&&+l_{*}^2 Q_{*}^4 \left[3 \left(5 e^2-4\right) Q_{*}^2+4 \left(2 e^4-3 e^2+1\right) Q_{*}^4+8\right] \nonumber\\
&&+Q_{*}^8 +\left(e^2-1\right) Q_{*}^{10}=0.
\end{eqnarray}
We can obtain its analytic form but we do not show it since it is a very complicated and we do not use the general case. 

Here, we concentrate on marginal energy case $E=m$ or $e=1$. 
In the case, Eq.~(\ref{eq:lcRN1}) yields  
\begin{eqnarray}\label{eq:lcRN2}
&&l_*^6 \left(Q_*^2-1\right)+l_*^4 \left(3 Q_*^4-20 Q_*^2+16\right) \nonumber\\
&&+l_*^2 \left(3 Q_*^2+8\right) Q_*^4+Q_*^8 =0.
\end{eqnarray}
and its maximum root gives $l=l_\mathrm{*cRN}$.
We show the analytic form of $l_\mathrm{*cRN}$ as
\begin{eqnarray}
l_\mathrm{*cRN} 
&=& 
\left(\frac{1}{6 \left(Q_{*}^2-1\right) H} \right)^\frac{1}{2} 
\left[ 270 \sqrt[3]{-2} Q_{*}^6+ (-2)^{2/3} H^2 \right. \nonumber\\
&&-32 \left(H+16 \sqrt[3]{-2}\right) +40 Q_{*}^2 \left( H+32 \sqrt[3]{-2}\right) \nonumber\\
&&\left. -2 Q_{*}^4 \left(3 H+520 \sqrt[3]{-2}\right) \right]^\frac{1}{2},
\end{eqnarray}
where $H$ is 
\begin{eqnarray}
H&\equiv& \left( 729 Q_{*}^{10}-8775 Q_{*}^8+29680 Q_{*}^6-44160 Q_{*}^4+30720 Q_{*}^2 \right. \nonumber\\
&&\left. +3 \sqrt{3} \sqrt{Q_{*}^{10} \left(Q_{*}^2-1\right)^2 \left(27 Q_{*}^2-32\right)^3}-8192 \right)^\frac{1}{3}
\end{eqnarray}
but it is still complicated.
Given the value of $Q_*$, 
we can solve Eq.~(\ref{eq:lcRN2}) easily in numerical and analytical expressions to obtain $l_\mathrm{*cRN}$.

\section{Comparison of the center-of-mass energy with $Q/M=q/M$}
The center-of-mass energy~(\ref{eq:Ecm1})
in the GMGHS spacetime with $q=M(1-\epsilon_q)$, where $0<\epsilon_q \ll 1$, 
can be expanded as
\begin{eqnarray}\label{eq:Ecm1e}
\frac{E_\mathrm{CM}(r_\mathrm{H})}{2m} 
&=&\frac{1}{\sqrt{2}}\epsilon_q^{-1/2}+1+\frac{5}{4 \sqrt{2}}\epsilon_q^{1/2}-\epsilon_q+\frac{43}{32 \sqrt{2}}\epsilon_q^{3/2} \nonumber\\
&&-\frac{335}{128 \sqrt{2}}\epsilon_q^{5/2}+O(\epsilon_q ^3). 
\end{eqnarray}
The center-of-mass energy~(\ref{eq:Ecm2})
in the RN spacetime with $Q_*=M(1-\epsilon_Q)$ 
can be expanded as 
\begin{eqnarray}\label{eq:Ecm2e}
\frac{E_\mathrm{CM}(r_\mathrm{H})}{2m} 
&=&\sqrt{\frac{1}{2} \left(5 \sqrt{5}+13\right)} 
+\frac{\left(-5 \sqrt{5}-11\right)}{\sqrt{5 \sqrt{5}+13}}\epsilon_Q^{1/2} \nonumber\\
&&+\frac{2 \sqrt{2} \left(81 \sqrt{5}+181\right) }{\left(5 \sqrt{5}+13\right)^{3/2}} \epsilon_Q\nonumber\\
&&-\frac{\left(7275 \sqrt{5}+16267\right) }{\left(5 \sqrt{5}+13\right)^{5/2}}\epsilon_Q^{3/2}\nonumber\\
&&+\frac{8 \sqrt{2} \left(66986 \sqrt{5}+149785\right)}{5 \left(5 \sqrt{5}+13\right)^{7/2}} \epsilon_Q^2\nonumber\\
&&-\frac{\left(3692855 \sqrt{5}+8257474\right) }{\left(5 \sqrt{5}+13\right)^{9/2}}\epsilon_Q ^{5/2} +O(\epsilon_Q^3).  \nonumber\\
\end{eqnarray}

We assume $Q/M=q/M$ and we introduce $\epsilon \equiv \epsilon_q=\epsilon_Q$.
We can express $\mathrm{Eq.}~(\ref{eq:Ecm1e}) - \mathrm{Eq.}~(\ref{eq:Ecm2e})$
and $\mathrm{Eq.}~(\ref{eq:Ecm1e})/\mathrm{Eq.}~(\ref{eq:Ecm2e})$ as
\begin{eqnarray}\label{eq:Ecmdif}
&&\mathrm{Eq.}~(\ref{eq:Ecm1e}) - \mathrm{Eq.}~(\ref{eq:Ecm2e}) \nonumber\\
&=&\frac{1}{\sqrt{2}}\epsilon^{-1/2}+\left(1-\sqrt{\frac{1}{2} \left(5 \sqrt{5}+13\right)}\right) \nonumber\\
&&+\left(\frac{5 \sqrt{5}+11}{\sqrt{5 \sqrt{5}+13}}+\frac{5}{4 \sqrt{2}}\right) \epsilon^{1/2} \nonumber\\
&&-\left(1+\frac{2 \sqrt{2} \left(81 \sqrt{5}+181\right)}{\left(5 \sqrt{5}+13\right)^{3/2}}\right) \epsilon +O(\epsilon^{3/2}) 
\end{eqnarray}
and 
\begin{eqnarray}\label{eq:Ecmrat}
\frac{\mathrm{Eq.}~(\ref{eq:Ecm1e})}{\mathrm{Eq.}~(\ref{eq:Ecm2e})} 
&=&\frac{1}{\sqrt{5 \sqrt{5}+13}}\epsilon^{-1/2} 
+\frac{2 \left(12 \sqrt{2}+5 \sqrt{10}\right)}{\left(5 \sqrt{5}+13\right)^{3/2}} \nonumber\\
&&+\frac{\left(597 \sqrt{5}+1343\right)}{2 \left(5 \sqrt{5}+13\right)^{5/2}} \epsilon^{1/2} \nonumber\\
&&-\frac{16 \left(348213 \sqrt{2}+155725 \sqrt{10}\right)}{\left(5 \sqrt{5}+13\right)^{11/2}} \epsilon \nonumber\\
&&+O(\epsilon^{3/2}),
\end{eqnarray}
respectively,
and we show numerical values in Table~I.
Figure 7 shows the ratio of $E_\mathrm{CM}(r_\mathrm{H})/(2m)$ given by Eq.~(\ref{eq:Ecm1}) in the GMGHS spacetime to Eq.~(\ref{eq:Ecm2}) in the RN spacetime. 
\begin{figure}[htbp]
\begin{center}
\includegraphics[width=80mm]{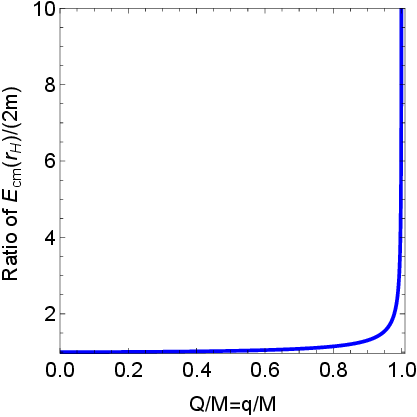}
\end{center}
\caption{
\label{fig:EcmRatio} 
The ratio of $E_\mathrm{CM}(r_\mathrm{H})/(2m)$ given by Eq.~(\ref{eq:Ecm1}) in the GMGHS spacetime to Eq.~(\ref{eq:Ecm2}) in the RN spacetime. 
We have set $Q/M=q/M$ to compare them.
}
\end{figure}
\begin{table}[htbp]
 \caption{We assume $Q/M=q/M$ and we introduce $\epsilon \equiv \epsilon_q=\epsilon_Q$.
 Eqs.~(\ref{eq:Ecm1}) and (\ref{eq:Ecm2}), Eq.~(\ref{eq:Ecm1})$-$Eq.~(\ref{eq:Ecm2}),  Eq.~(\ref{eq:Ecm1})$/$Eq.~(\ref{eq:Ecm2}), and Eqs.~(\ref{eq:Ecmdif}) and (\ref{eq:Ecmrat}) for given $\epsilon$ are shown in numerical.
 }
\begin{center}
\begin{tabular}{|c|c|c|c|c|} \hline
$\epsilon$                                         &$10^{-2}$ &$10^{-4}$ &$10^{-6}$ &$10^{-8}$ \\ \hline
Eq.~(\ref{eq:Ecm1})                                &$8.15039$ &$71.7194$ &$708.108$ &$7072.07$ \\ \hline
Eq.~(\ref{eq:Ecm2})                                &$3.10175$ &$3.43284$ &$3.47259$ &$3.47664$ \\ \hline
Eq.~(\ref{eq:Ecm1})$-$Eq.~(\ref{eq:Ecm2})          &$5.04864$ &$68.2866$ &$704.635$ &$7068.59$ \\ \hline
To $\epsilon^{-1/2}$ term in Eq.~(\ref{eq:Ecmdif}) &$7.07107$ &$70.7107$ &$707.107$ &$7071.07$ \\ \hline
To $\epsilon^0$ term in Eq.~(\ref{eq:Ecmdif})      &$4.59398$ &$68.2336$ &$704.630$ &$7068.59$ \\ \hline
To $\epsilon^{1/2}$ term in Eq.~(\ref{eq:Ecmdif})  &$5.13343$ &$68.2875$ &$704.635$ &$7068.59$ \\ \hline
To $\epsilon^1$ term in Eq.~(\ref{eq:Ecmdif})      &$5.03729$ &$68.2866$ &$704.635$ &$7068.59$ \\ \hline
Eq.~(\ref{eq:Ecm1})$/$Eq.~(\ref{eq:Ecm2})          &$2.62767$ &$20.8922$ &$203.913$ &$2034.17$ \\ \hline
To $\epsilon^{-1/2}$ term in Eq.~(\ref{eq:Ecmrat}) &$2.03362$ &$20.3362$ &$203.362$ &$2033.62$ \\ \hline
To $\epsilon^0$ term in Eq.~(\ref{eq:Ecmrat})      &$2.58502$ &$20.8876$ &$203.913$ &$2034.17$ \\ \hline
To $\epsilon^{1/2}$ term in Eq.~(\ref{eq:Ecmrat})  &$2.63159$ &$20.8922$ &$203.913$ &$2034.17$ \\ \hline
To $\epsilon^1$ term in Eq.~(\ref{eq:Ecmrat})      &$2.62771$ &$20.8922$ &$203.913$ &$2034.17$ \\ \hline
\end{tabular}
\end{center}
\end{table}




\end{document}